\documentclass{PoS}

\usepackage{lineno}
\usepackage{cite}
\usepackage[htt]{hyphenat}
\usepackage[center]{subfigure}
\usepackage{url}

\title{Determination of proton parton distribution functions using ATLAS data}

\ShortTitle{Constraining PDFs using ATLAS data}

\author{\speaker{Francesco Giuli}\thanks{on behalf of the ATLAS Collaboration}\\
        \\University of Rome Tor Vergata and INFN, Sezione di Roma 2, Via della Ricerca Scientifica 1, 00133 Roma, Italy\\
        E-mail: \email{francesco.giuli@roma2.infn.it}}

\abstract{Fits to determine parton distribution functions using top-antitop, inclusive $W/Z$ boson and $W^{\pm}$ boson production measurements in association with jets from ATLAS, in combination with deep-inelastic scattering data from HERA, are presented. The ATLAS $W/Z$ boson data exhibit sensitivity to the valence quark distributions and the light quark sea composition, whereas the top-quark pair production data have sensitivity to the gluon distribution. The impact of the these data is increased by fitting several distributions simultaneously, with the full information on the systematic and statistical correlations between data points. The parton distribution functions extracted using $W^{\pm}$ + jets data show an improved determination of the high-$x$ sea-quark densities, while confirming the unsuppressed strange-quark density at lower $x$ < 0.02 found by previous ATLAS analyses.}

\FullConference{
European Physical Society Conference on High Energy Physics - EPS-HEP2019 -\\
			10-17 July, 2019\\
			Ghent, Belgium}

\begin{document}

\begin{figure}[t]
\centering
\includegraphics[width=7.51cm]{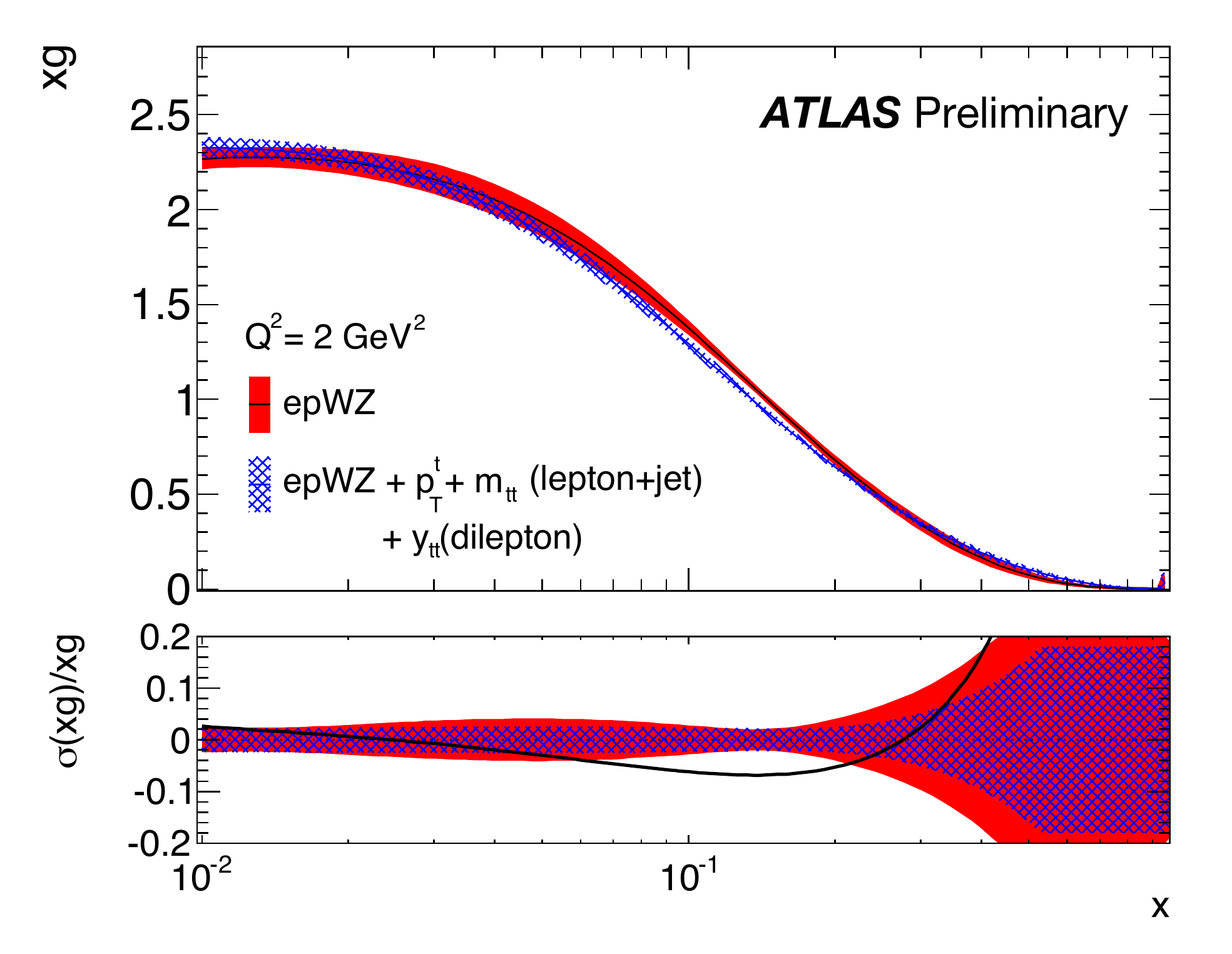}
\includegraphics[width=7.51cm]{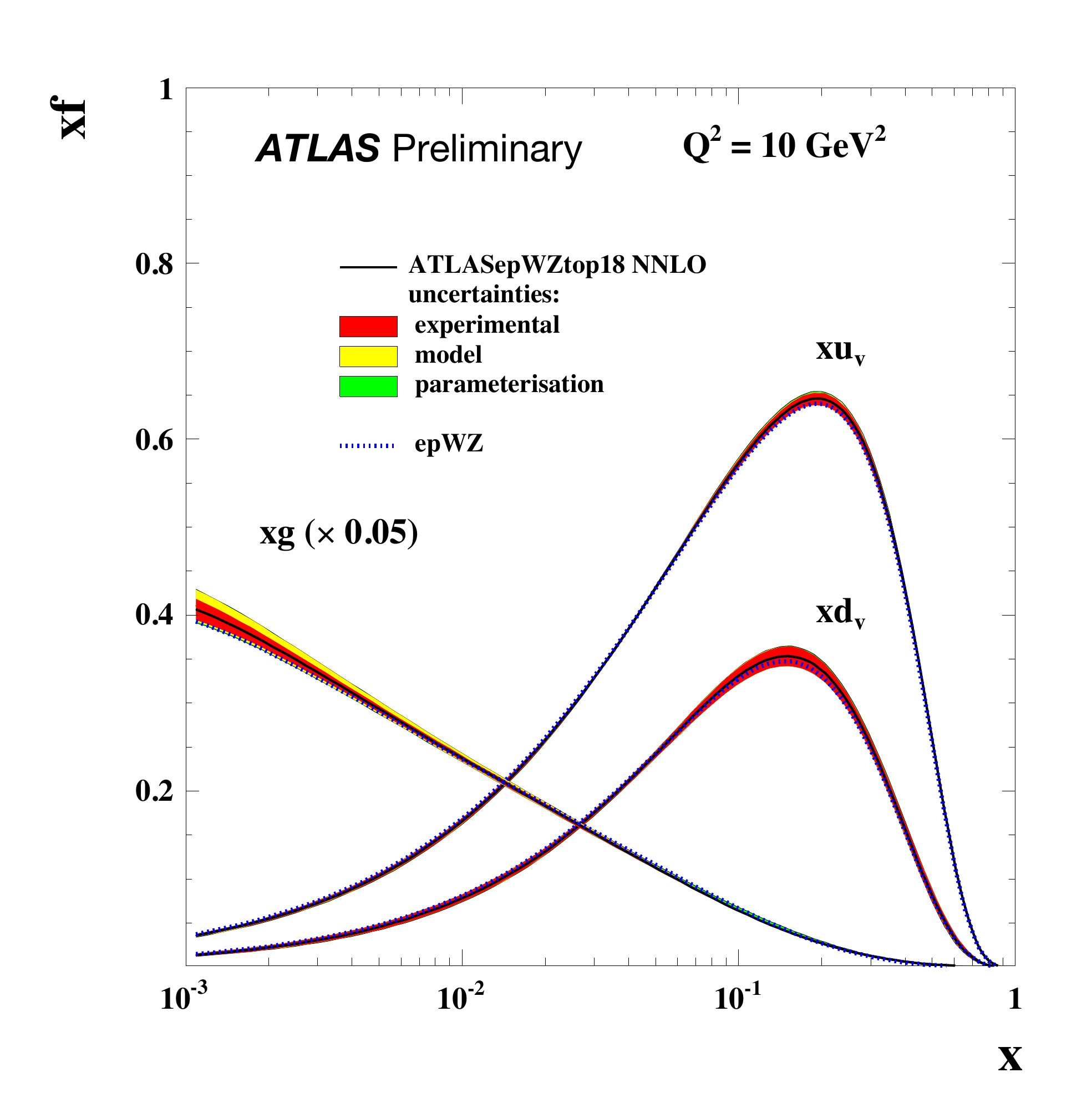}
\caption{Left: Gluon PDF from a fit to HERA and ATLAS $W$, $Z/\gamma^{*}$ boson data plus the lepton+jets $m_{t\bar{t}}$ and $p_{\mathrm{T}}^{t}$ and the dilepton $y_{t\bar{t}}$ compared to a fit to HERA and ATLAS $W$, $Z/\gamma^{*}$ boson data only. Only experimental uncertainties on the input data have been included. Right: The valence and gluon PDFs from a fit to HERA and ATLAS $W$, $Z/\gamma^{*}$ boson data plus the lepton+jets $m_{t\bar{t}}$ and $p_{\mathrm{T}}^{t}$ and the dilepton $y_{t\bar{t}}$ are shown, including model and parametrisation uncertainties. Plots taken from Ref.~\cite{ATL-PHYS-PUB-2018-017}.}
\label{fig:gluon_top}
\end{figure}

\paragraph{Introduction}
A key ingredient of physics with hadrons in the initial state are the Parton Distribution Functions (PDFs), which describes the longitudinal momentum fraction $x$ carried out by partons in a proton. Thus, PDFs represent a fundamental aspect of perturbative QCD phenomenology.
High-precision measurements of Standard Model (SM) processes can be used to put constraints on PDFs, allowing comparisons with the current precision reached theoretically. In this document, two different QCD analyses are presented. The first presents the impact of including ATLAS top-quark pair ($t\bar{t}$) production data~\cite{top_8ljets,top_dilep}, while the second evaluates the impact of including ATLAS $W^{+}$ and $W^{-}$ boson production measurements in association with jets~\cite{Wjets_ATLAS}.\\
In both these analyses, data are studied in combination with the final neutral-current (NC) and charged-current (CC) deep inelastic scattering (DIS) HERA I+II data~\cite{Abramowicz2015} and the ATLAS precision measurements of the inclusive differential $W$ and $Z/\gamma^{*}$ boson cross section data at $\sqrt{s}$ = 7 TeV~\cite{ATLAS_WZ7TeV}.
\vspace{-3.55mm}
\paragraph{Impact of $t\bar{t}$ cross sections data on PDFs}
These data are complementary to the $W$ and $Z/\gamma^{*}$ boson data in their PDF constraining power since they are sensitive to the high-$x$ gluon distribution ($x\gtrsim$ 0.05). They have been measured at 8 TeV using 20.2 fb$^{-1}$ of data in the lepton+jets~\cite{top_8ljets} and dilepton~\cite{top_dilep} decay modes. While the available spectra in the lepton+jets channel are the invariant mass of the $t\bar{t}$ system, $m_{t\bar{t}}$, the rapidity of the $t\bar{t}$ pair, $y_{t\bar{t}}$, the average top-quark rapidity, $y_{t}^{\mathrm{avg}}$ and the top-quark transverse momentum, $p_{\mathrm{T}}^{t}$, only the $m_{t\bar{t}}$ and $y_{t\bar{t}}$ spectra are available in the dilepton channel. In the lepton+jets channel,  all the available differential spectra have full information on systematic bin-to-bin correlations and the 55 sources of systematic uncertainties are correlated among the different spectra. For the dilepton data the correlations are provided as a total covariance matrix for each spectrum separately, so that only one of these distributions may be fitted at a time.\\
The {\tt{xFitter}} framework~\cite{xFitter} is used for the present QCD analysis, where the PDF evolution is performed using DGLAP at NNLO through {\tt{QCDNUM}}~\cite{qcdnum}. Calculations for the \textit{W} and $Z/\gamma^{*}$ boson production are made at NNLO in QCD and NLO in electroweak (EW), while the NNLO predictions of perturbative QCD for top-quark pair production data have recently become available~\cite{top_NNLO}.\\
The initial scale $Q_0^{2}$ where PDFs are parametrised as a function of $x$ is chosen to be 1.9 GeV$^2$, such that it is below the charm mass threshold $m_c^{2}$. The heavy quark masses are chosen to be $m_c$ = 1.43 GeV and $m_b$ = 4.5 GeV. The strong coupling constant is fixed to $\alpha_{s}(M_{Z})$ = 0.118. A minimum cut of $Q^{2}_{\mathrm{min}}\geq$ 10.0 GeV$^2$ is imposed on the HERA data. All these assumptions are varied in the evaluation of model uncertainties on the final fit.\\
The functional form used for parametrising PDFs is defined as follows:
\begin{equation}
\label{eq:PDFpara}
xq_{i}(x) = A_{i}x^{B_{i}}(1-x)^{C_{i}}P_{i}(x)
\end{equation}
where $i$ represents the flavour of the quark distribution, $P_{i}(x) = (1+D_{i}x+E_{i}x^{2})e^{F_{i}x}$ and the $A_{i}-F_{i}$ coefficients are the parameters which are minimized in the fit. The $B$-parameters determine the low-$x$ behaviour and $C$-parameters regulate the high-$x$ regime. The chosen PDFs to be  parametrised at the starting scale are the gluon distribution, the valence quark distributions, $xu_{v}$ and $xd_{v}$, and the light anti-quark distributions, $x\bar{u}, x\bar{d}$ and $x\bar{s}$. A more flexible form, namely $xg(x)=A_{g}x^{B_{g}}(1-x)^{C_{g}}P_{g}(x)-A'_{g}x^{B'_{g}}(1-x)^{C'_{g}}$ (with $C'_{g} =$ 25 to suppress negative contributions at high-$x$) is used for parametrising the gluon PDF. The $\chi^{2}$ used to compare experimental measurements, $\mu_{i}$, with theoretical predictions, $m_{i}$, is defined as:
\begin{equation}
\label{eq:chi2DEF}
\chi^{2}=\sum_{ij}\left(m_{i}-\sum_{k}\gamma_{ki}b_{k}-\mu_{i}\right)C^{-1}_{\mathrm{stat\; ij}}\left(m_{j}-\sum_{k}\gamma_{kj}b_{k}-\mu_{j}\right)+\sum_{k}b_{k}^{2}
\end{equation}
where $C_{\mathrm{stat\; ij}}$ represents the statistical correlations between data points $i,\;j$ and the systematic correlations are treated by nuisance parameters, $b_{k}$, for each source of systematic uncertainty $k$. The 1$\sigma$ deviation correlated systematic uncertainty on point $i$ due to the systematic uncertainty $k$ is represented by the $\gamma_{ki}$ quantities, while the $\sum_{k}b_{k}^{2}$ term takes the $\pm1\sigma$ constraints of the nuisance parameters into account.\\
A fit to the $m_{t\bar{t}}$ and $p_{\mathrm{T}}^{t}$ spectra from lepton+jets and the $y_{t\bar{t}}$ spectrum from the dilepton data has been performed, and it is called the ATLASepWZtop18 PDF fit. Table~\ref{Chi2_top} shows the $\chi^{2}$ for this fit.
\begin{table}
\centering
\begin{tabular}{lcc}
\hline
 & & lepton+jets $p^t_{\mathrm{T}}$, $m_{t\bar{t}}$ \\
& &and  dilepton $y_{t\bar{t}}$ spectra \\
\hline
  Total $\chi^2/\rm{NDF}$ &                                  & 1253.8 / 1061        \\
  Partial $\chi^2/\rm{NDP}$  &HERA                          & ~~1149 / 1016          \\
  Partial $\chi^2/\rm{NDP}$  &ATLAS $W,Z/\gamma^{*}$               & 78.9 / 55            \\
  Partial $\chi^2/\rm{NDP}$ & ATLAS lepton+jets $p^t_{\mathrm{T}}$, $m_{t\bar{t}}$&16.0 / 15            \\
   Partial $\chi^2/\rm{NDP}$             &ATLAS dilepton $y_{t\bar{t}}$ &5.4 / 5~              \\
\hline
\end{tabular}
\caption{Total and partial $\chi^{2}$ for datasets entering the PDF fit to dilepton $y_{t\bar{t}}$ spectrum and the lepton+jets $m_{t\bar{t}}$ and $p_{\mathrm{T}}^{t}$ spectra.}
\label{Chi2_top}
\end{table}
The gluon PDF distribution before and after the three above-mentioned $t\bar{t}$ spectra are added to the HERA and ATLAS $W$ and $Z/\gamma^{*}$ boson data. A significantly harder gluon for $x>0.1$ and a reduced uncertainty on the gluon PDF in the high-$x$ regime can be observed in Figure~\ref{fig:gluon_top} (left). Then, additional uncertainties due to possible model and parameterisation biases are evaluated. The model uncertainties include effects related to variations of the charm (1.37 $<m_{c}<$ 1.49 GeV), beauty (4.25 $<m_{b}<$ 4.75 GeV) and top (172.3 $<m_{t}<$ 175.0 GeV) quark masses, of the $Q^{2}_{\mathrm{mim}}$ cut on the data to enter the fit (7.5 $<Q^{2}_{\mathrm{mim}}<$ 12.5 GeV$^{2}$) and the value of the starting scale $Q_{0}^{2}$ (1.6 $<Q_{0}^{2}<$ 2.2 GeV$^{2}$). The parameterisation uncertainties correspond to the envelope of the results obtained with the following extra parameters added in the functions $P_{i}$ in Eq.~\ref{eq:PDFpara}: the $D_{g}$ parameter to the gluon PDF and freeing $B_{\bar{s}}$. Furthermore, the impact of adding both $E_{g}$ and $F_{g}$ term to the gluon has been studied, and it does not bring any additional modification to the gluon shape. The valence and gluon PDFs with model and parametrisation uncertainties are shown in Figure~\ref{fig:gluon_top} (right).

\begin{figure}[t]
\centering
\includegraphics[width=7.51cm]{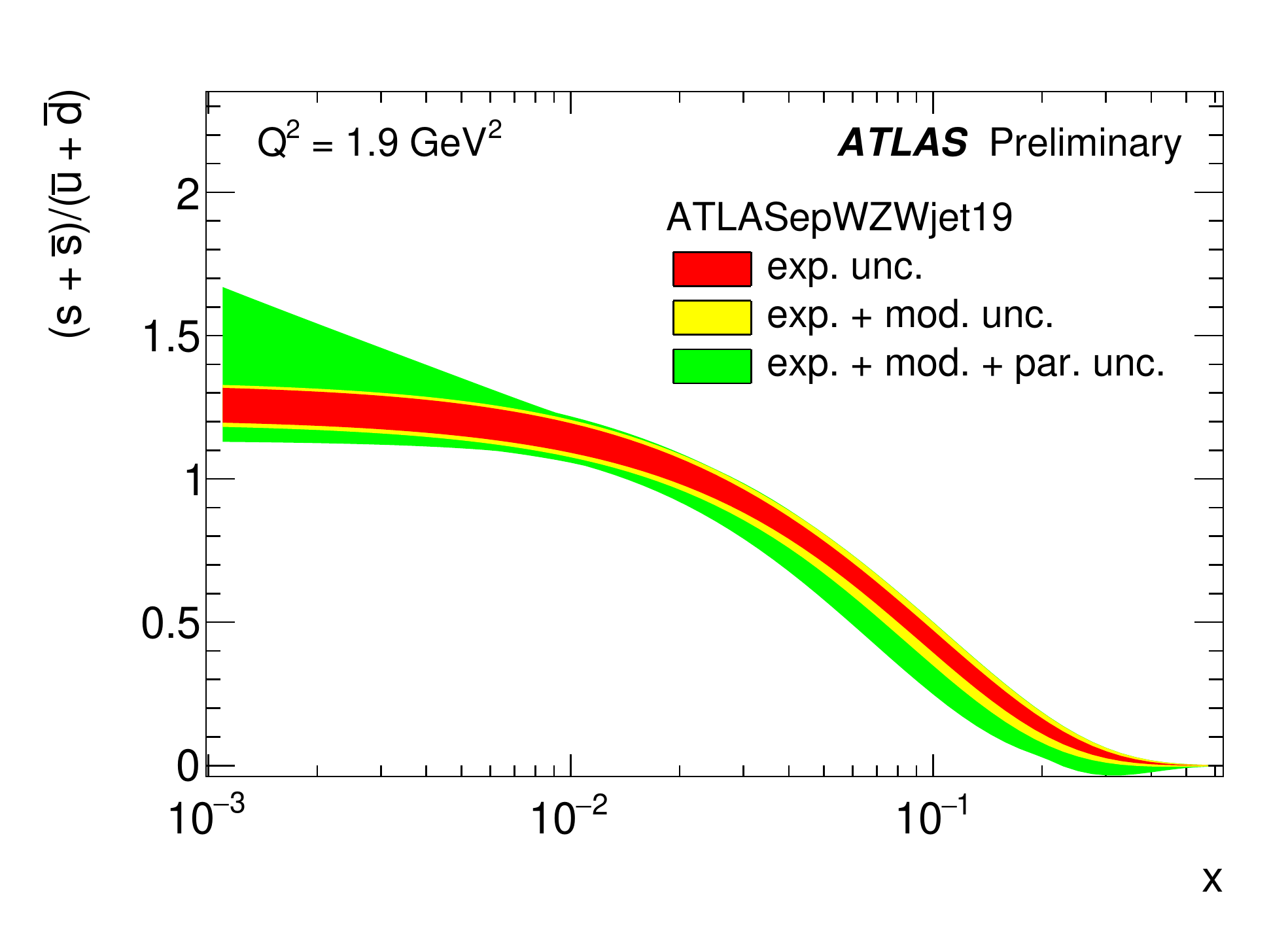}
\includegraphics[width=7.51cm]{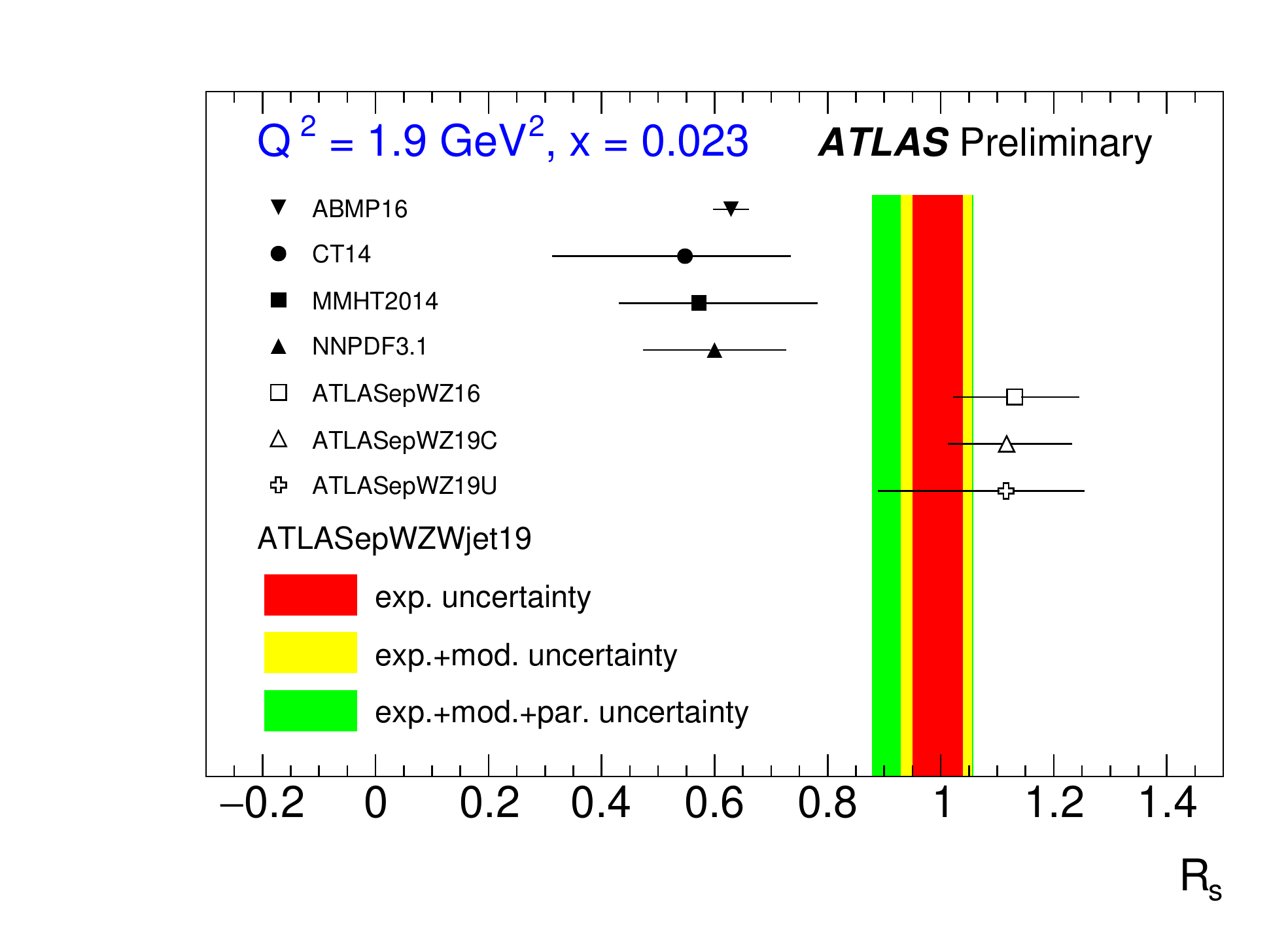}
\caption{Left: $R_{s}$ distribution, evaluated at $Q^{2}$ = 1.9 GeV$^{2}$, predicted by the ATLASepWZWjet19 fit. Right: $R_{s}$ evaluated at $x$ = 0.023 and $Q^{2}$ = 1.9 GeV$^{2}$ for the ATLASepWZWjet19 PDF set in comparison to global PDFs~\cite{CTEQ,MMHT,NNPDF,ABMP}, and the ATLASepWZ16, ATLASepWZ19U and ATLASepWZ16C sets. The experimental, model and parameterisation uncertainty bands are plotted separately for the ATLASepWZWjet19 results. All uncertainty bands are at 68\% confidence level. Plots taken from Ref.~\cite{ATL-PHYS-PUB-2019-016}.}
\label{fig:Rs}
\end{figure}

\begin{figure}[t]
\centering
\includegraphics[width=7.51cm]{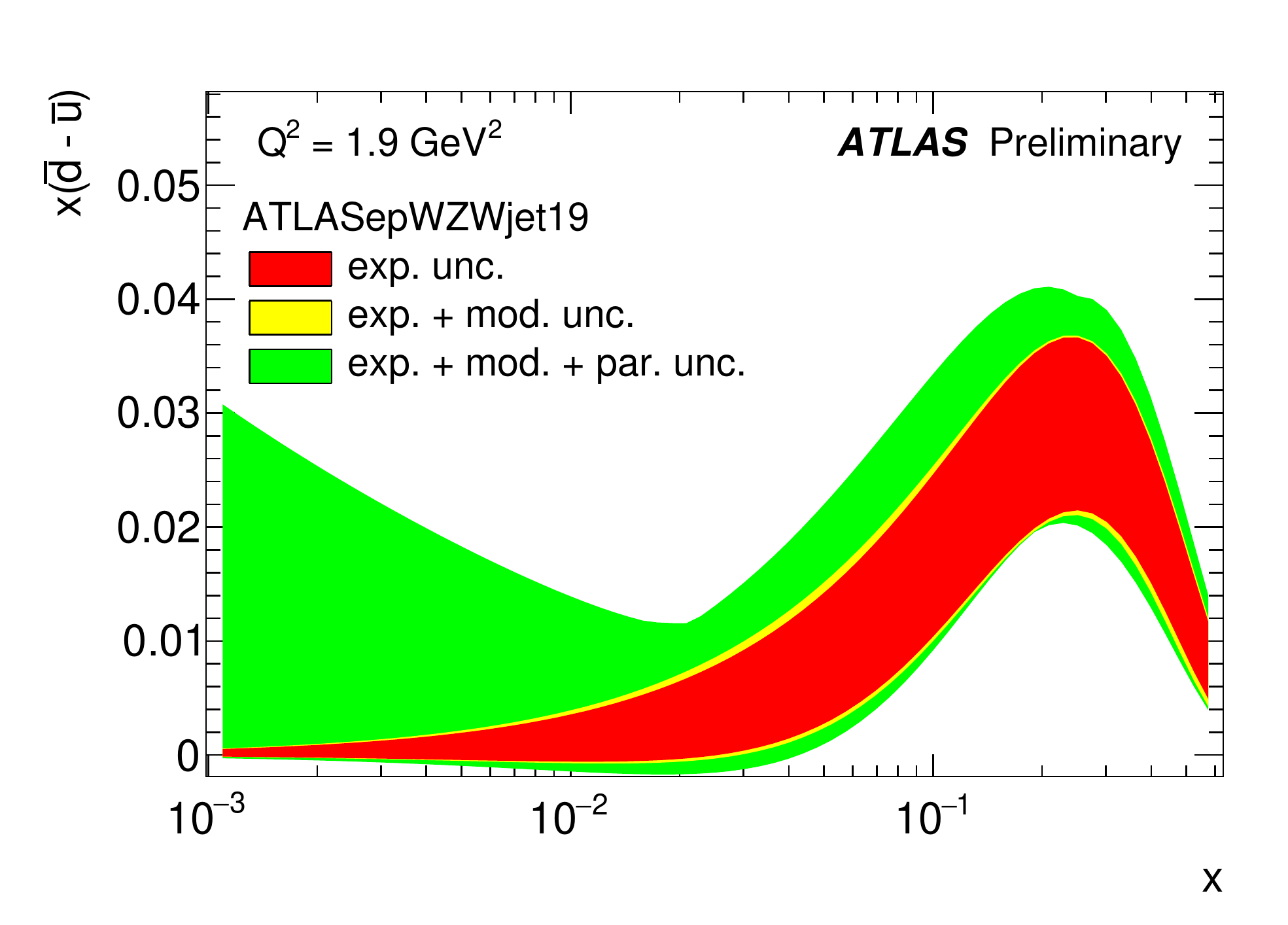}
\includegraphics[width=7.51cm]{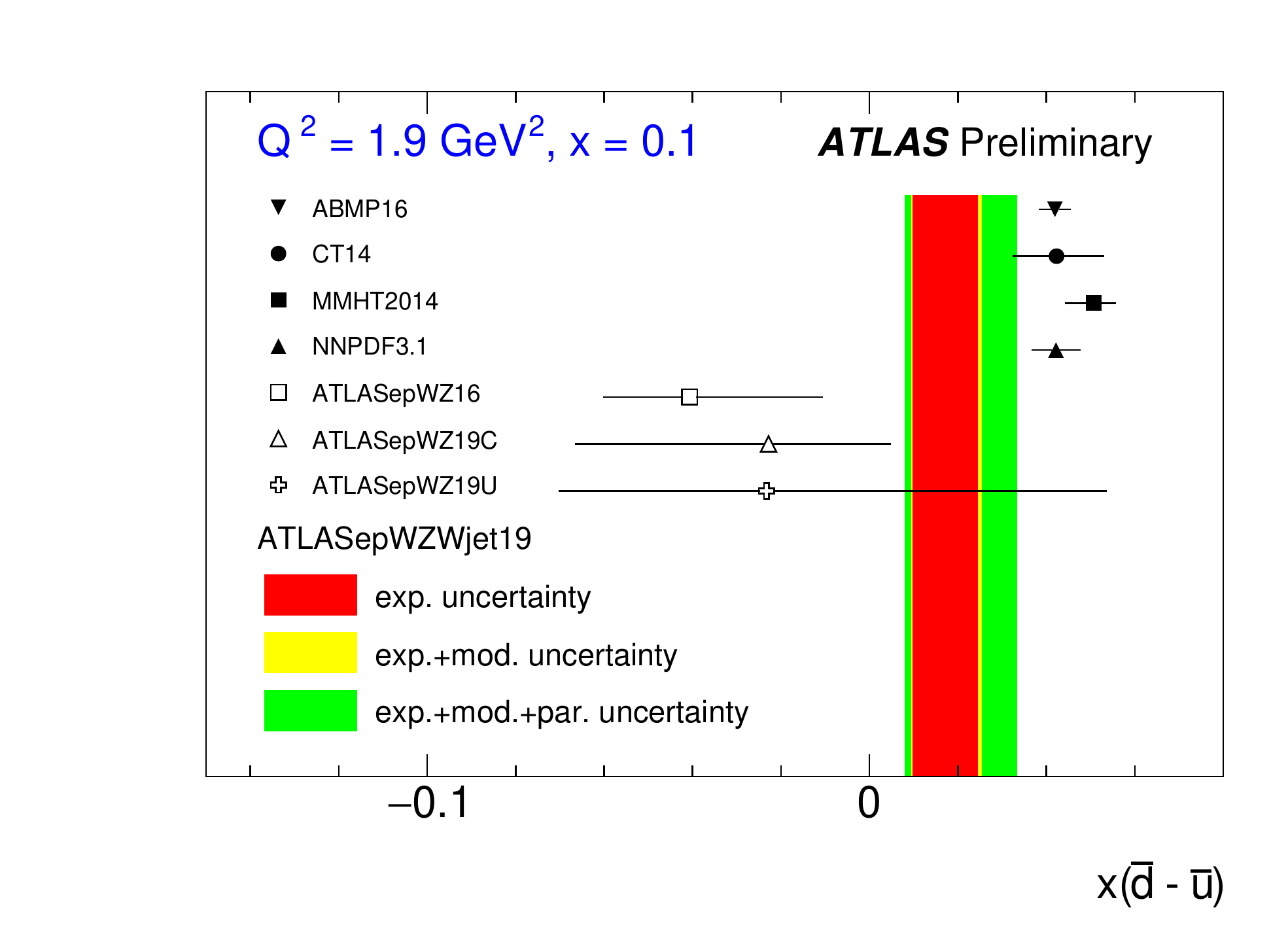}
\caption{Left: $x(\bar{d}-\bar{u})$ distribution, evaluated at $Q^{2}$ = 1.9 GeV$^{2}$, predicted by the ATLASepWZWjet19 fit. Right: $x(\bar{d}-\bar{u})$ evaluated at $x$ = 0.1 and $Q^{2}$ = 1.9 GeV$^{2}$ for the ATLASepWZWjet19 PDF set in comparison to global PDFs~\cite{CTEQ,MMHT,NNPDF,ABMP}, and the ATLASepWZ16, ATLASepWZ19U and ATLASepWZ16C sets. The experimental, model and parameterisation uncertainty bands are plotted separately for the ATLASepWZWjet19 results. All uncertainty bands are at 68\% confidence level. Plots taken from Ref.~\cite{ATL-PHYS-PUB-2019-016}.}
\label{fig:uBAR}
\end{figure}

\paragraph{QCD analysis of ATLAS $W^{\pm}$ boson production data in association with jets}
The ATLAS $W^{\pm}$ + jets differential cross sections are based on data recorded in $pp$ collisions at $\sqrt{s}$ = 8 TeV, with a total integrated luminosity of 20.2 fb$^{-1}$ and, they are provided in the electron channel only. These data are divided in $W^+$ and $W^-$ cross sections, and correlations among them are fully considered. Four differential spectra are available: the transverse momentum of the $W$ boson, $p_{\mathrm{T}}^W$, the transverse momentum of the leading jet, $p_{\mathrm{T}}^{\mathrm{leading}}$, the absolute rapidity of the leading jet, $|y^{\mathrm{leading}}|$ and the scalar sum of transverse momenta of the electron, all jets with $p_{\mathrm{T}}>$ 30 GeV and the missing transverse momentum ($E_{\mathrm{T}}^{\mathrm{miss}}$) in the event, $H_{\mathrm{T}}$. There are 50 sources of correlated systematic uncertainty in common between the different spectra, as well as five sources of uncorrelated systematics, including the data statistics. Moreover, full information on the statistical bin-to-bin correlations in data is available for each spectrum.\\
The same {\tt{xFitter}} fitting framework used for the $t\bar{t}$ cross section analysis is also used to perform the present QCD analysis. Predictions for $W^{\pm}$ + jets are obtained at fixed order (FO) up to NNLO in QCD and to LO in EW using the $\mathtt{N_{\mathrm{jetti}}}$ program~\cite{Njetti}. Outputs from the {\tt{APPLGRID}} code~\cite{APPLgrid} are used for fast calculations at NLO in QCD and LO in EW, and $K$-factors to match the $\mathtt{N_{\mathrm{jetti}}}$ predictions are used. The choice of the theoretical input parameters and of the functional form to parametrised PDFs is the same as described in the previous section. A bias correction term, referred to as \textit{log penalty} term, is added to the $\chi^{2}$ formula in Eq.~\ref{eq:chi2DEF}, namely:
\begin{equation}
\label{eq:chi2_log}
\sum_{i}\log\frac{\delta_{i,\mathrm{uncor}}^{2}T_{i}^{2}+\delta_{i,\mathrm{stat}}^{2}D_{i}T_{i}}{\delta_{i,\mathrm{uncor}}^{2}D_{i}^{2}+\delta_{i,\mathrm{stat}}^{2}D_{i}^{2}}
\end{equation}
where $D_{i}$ represent the measured data, $T_{i}$ the corresponding theoretical predictions, $\delta_{i,\mathrm{uncor}}$ and $\delta_{i,\mathrm{stat}}$ are the uncorrelated systematic and the statistical uncertainties on $D_{i}$. In the following, we will refer to the first term in Eq.~\ref{Chi2_top} as \textit{partial} $\chi^{2}$, and to the second term as \textit{correlated} $\chi^{2}$. The statistical correlations considered in this analysis are the bin-to-bin correlations in $W^{\pm}$ + jets data, while there are no statistical correlations considered for the HERA and ATLAS $W$ and $Z/\gamma^{*}$ data.\\
A QCD analysis is then performed, fitting two different $W^{\pm}$ + jets individually, added on top of the HERA and ATLAS $W$ and $Z/\gamma^{*}$ data. Fitting more than one $W^{\pm}$ + jets spectrum at a time is not possible, as the spectra are highly correlated to each other, and information on the statistical correlation among them is not available. Table~\ref{Chi2_Wjets} shows the total and partial $\chi^2$ per degree of freedom (NDF) for each fit to the $W^{\pm}$ + jets spectra, as well as the correlated component of the $\chi^{2}$ and the log penalty term.
\begin{table}[t!]
\centering
\resizebox{\textwidth}{!}{%
\begin{tabular}{l c c c}
 \hline
 \hline
 Fit & ATLASepWZ19U & ATLASepWZ19U + $p_{\mathrm{T}}^{W}$  & ATLASepWZ19U + $p_{\mathrm{T}}^{\mathrm{leading}}$\\
 \hline
 Total $\chi^2/\mathrm{NDF}$ & 1310 / 1104 & 1354 / 1138 & 1365 / 1150\\
 HERA partial $\chi^2/\mathrm{NDF}$ & 1123 / 1016 & 1132 / 1016 & 1141 / 1016 \\
 HERA correlated $\chi^2$ & 48 & 49 & 50 \\
 HERA log penalty $\chi^2$ & -18 & -22 & -25 \\
 ATLAS $W$, $Z/\gamma^{*}$ partial $\chi^2/\mathrm{NDF}$ & 117 / 104 & 116 / 104 & 109 / 104 \\
 ATLAS $W^{\pm}$ + jets partial $\chi^2/\mathrm{NDF}$ & - & 18 / 34 & 43 / 46  \\
 ATLAS correlated $\chi^2$ & 40  & 62  & 47 \\
 ATLAS log penalty $\chi^2$ & -1 & -1 & 0 \\
 \hline
 \hline
\end{tabular}}
\caption{Total and partial $\chi^{2}$ for datasets entering the PDF fit, for each $W^{\pm}$ + jets separately, and the ATLASepWZ19U fit.}
\label{Chi2_Wjets}
\end{table}
\begin{figure}[t]
\centering
\includegraphics[width=4.95cm]{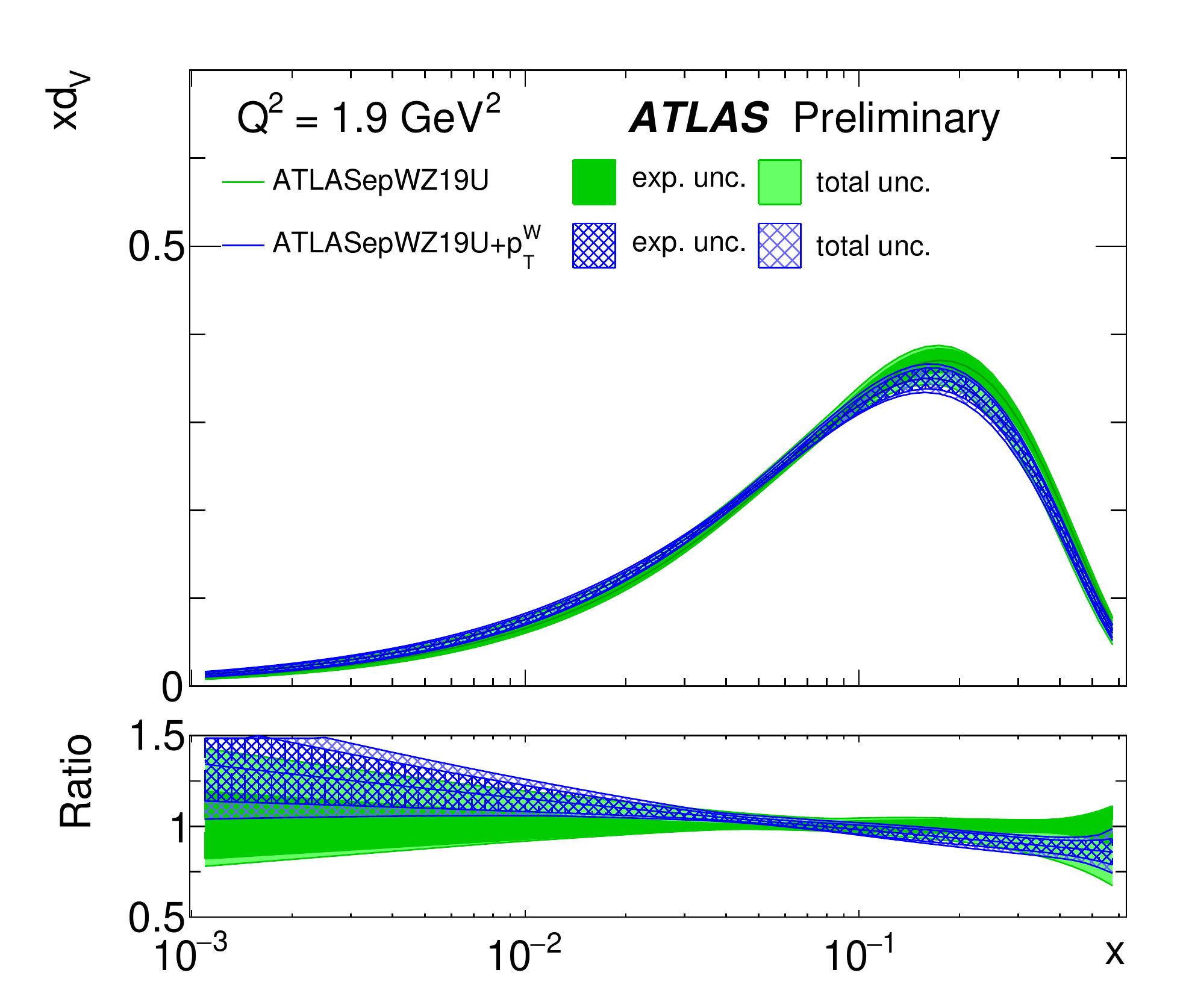}
\includegraphics[width=4.95cm]{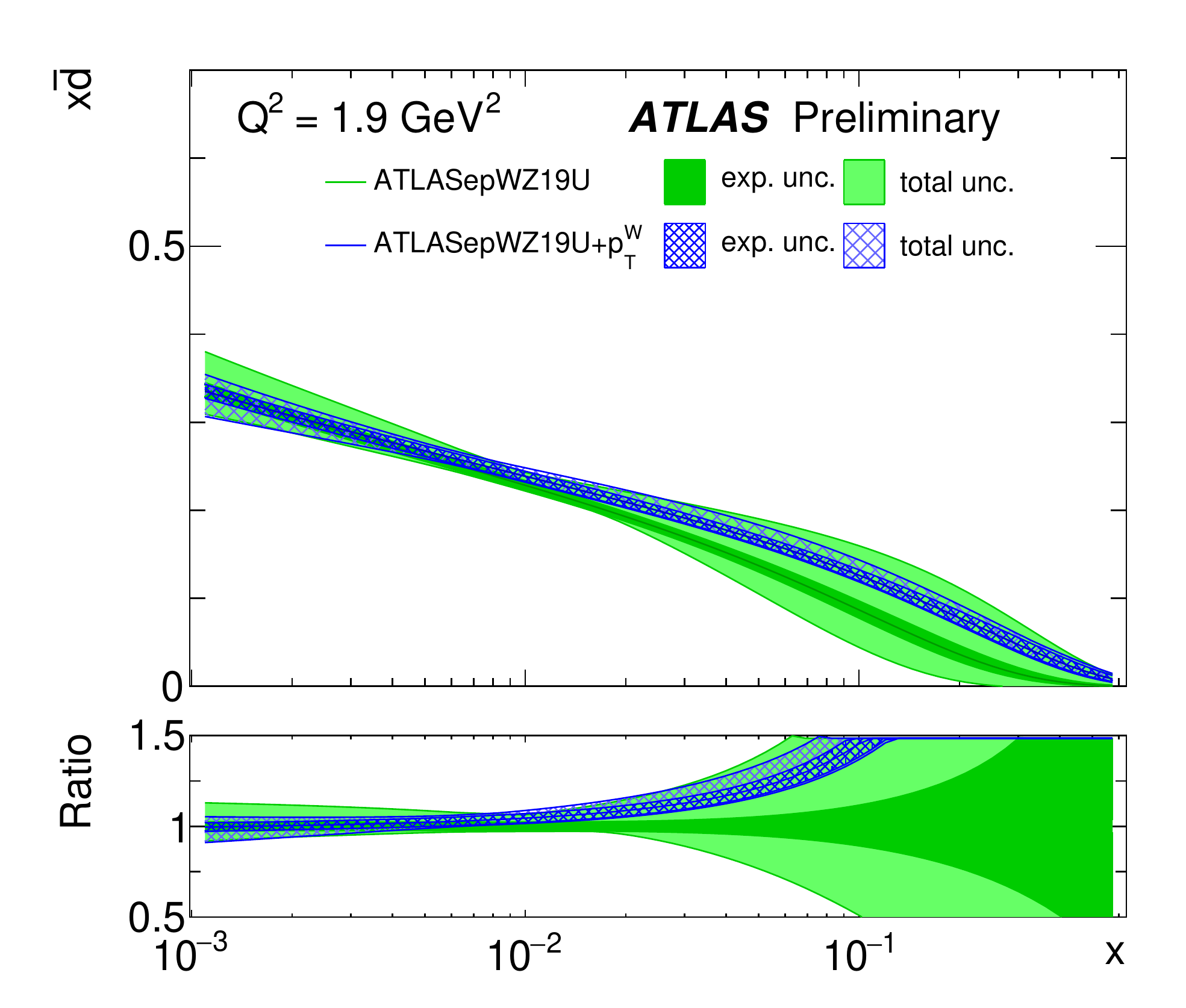}
\includegraphics[width=4.95cm]{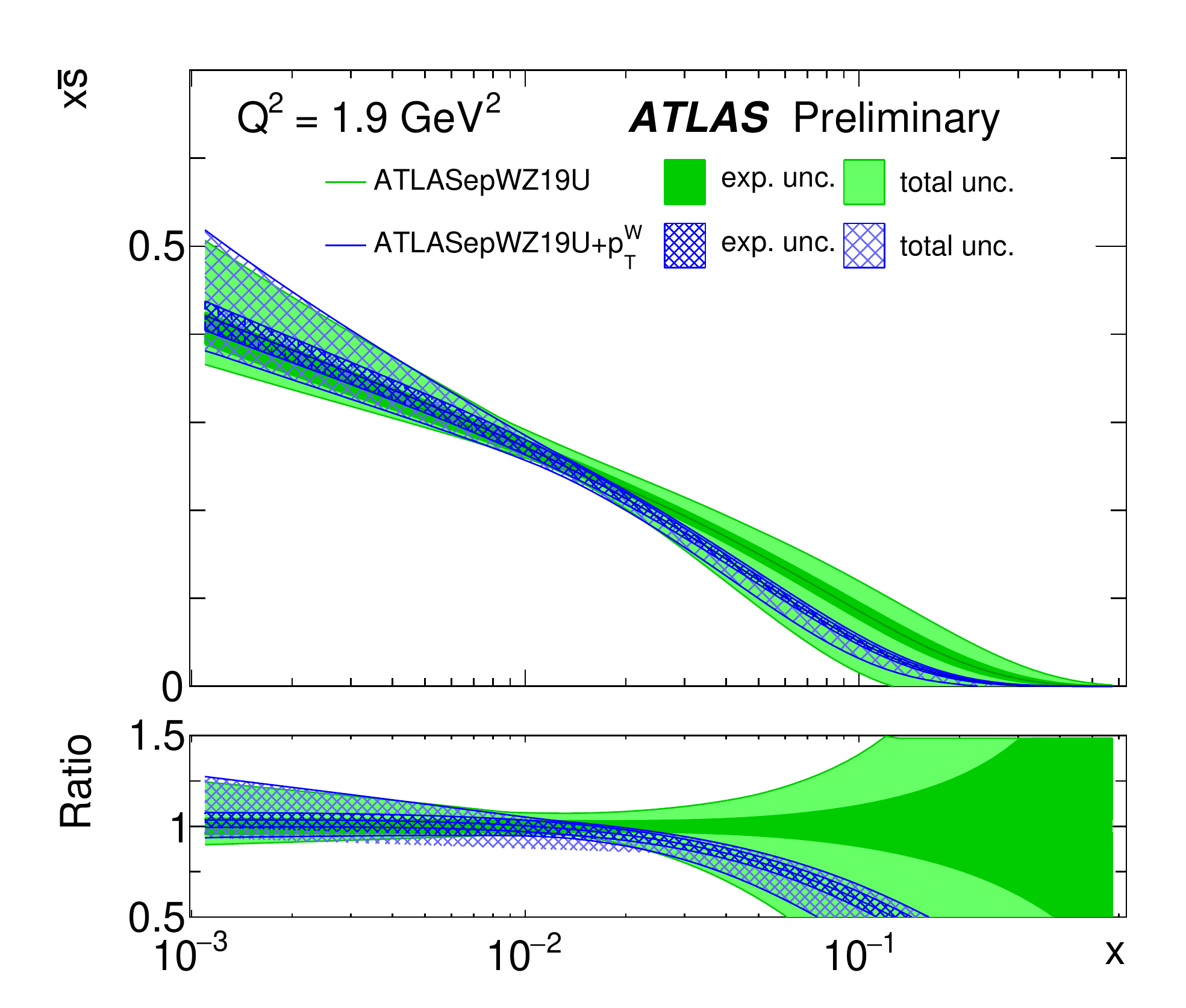}
\caption{Left: $xd_{v}$, Middle: $x\bar{d}$, Right: $x\bar{s}$ PDF obtained when fitting $W^{\pm}$ + jets, inclusive HERA and ATLAS $W$ and $Z/\gamma^{*}$ compared to a similar fit without the $W^{\pm}$ + jets data. Inner error bands indicate the experimental uncertainty, while outer bands the total uncertainty, which includes parameterisation and model uncertainties. Plots taken from Ref.~\cite{ATL-PHYS-PUB-2019-016}.}
\label{fig:Wjets_PDF}
\end{figure}
A good overall fit quality is visible, and the partial $\chi^{2}$ for the HERA and ATLAS $W$ and $Z/\gamma^{*}$ data are similar to those obtained in fits to the HERA+ATLAS $W$, $Z$ data alone. This demonstrates that there is no tension between these data and the $W^{\pm}$ + jets data. Note that the ATLASepWZ19U fit uses the electron and muon decay channels in the $W$ and $Z$ data separately. This choice has been made in order to correlate common sources of systematic uncertainties to those of the
$W^{\pm}$ + jets data. Furthermore, this fit has a different $Q^{2}_{\mathrm{min}}$ cutoff and a slightly different PDF parameterisation compared to the original ATLAS PDF fit where these data were included: the so-called ATLASepWZ16~\cite{ATLAS_WZ7TeV}.\\
The PDFs obtained from a fit with the $p_{\mathrm{T}}^{W}$ spectrum exhibit the smallest total uncertainty for all parton flavours, and the most sensitive distributions are shown in Figure~\ref{fig:Wjets_PDF} along with the ATLASepWZ19U fit for comparison. Here, the outer band represents the total uncertainty, which is the sum in quadrature of experimental, model and parameterisation uncertainties. Uncertainties due to model assumptions are evaluated following the same prescription highlighted in the previous section. The uncertainties related to the choice of the PDF parameterisation have been evaluated by adding additional parameters separately which give higher flexibility in the high-$x$ regions of all distributions. No additional parameters improve the fit $\chi^2$ by more than two, providing fits of similar quality. The resulting PDF set is named ``ATLASepWZWjet19''.\\
As regards the fraction of the strange-quark density in the proton can be characterised by the quantity $R_s$, defined by the ratio:
\begin{equation}\label{eq:Rs}
R_{s}=\frac{s+\bar{s}}{\bar{u}+\bar{d}}
\end{equation}
The $R_{s}$ distribution plotted as a function of $x$ evaluated at $Q^{2}$ = 1.9 GeV$^{2}$ is shown in Figure~\ref{fig:Rs} (left). The uncertainty bands are split in to the experimental, model and parameterisation uncertainties. For $x$ < 0.023, the fit with the $W^{\pm}$ + jets data maintains an unsuppressed strange-quark density, compatible with the previous result from the ATLASepWZ16 fit. The new ATLASepWZWjet19 fit favours a lower value of $R_{s}$ than the ATLASepWZ16 fit but is consistent to within 1$\sigma$. Tension remains with the global analyses~\cite{CTEQ,MMHT,NNPDF,ABMP} by more than one standard deviation in all cases, as shown in Figure~\ref{fig:Rs} (right).\\
Another interesting quantity to look at for judging the performance of PDF sets is the difference between $\bar{d}$ and $\bar{u}$ distributions. The $x(\bar{d}-\bar{u})$ as function of $x$ at $Q^{2}$ = 1.9 GeV$^{2}$ for the ATLASepWZWjets19 fit is shown in Figure~\ref{fig:uBAR} (left) with an overall positive $x(\bar{d}-\bar{u})$ distribution. This result is in contrast with what was previously found in the ATLASepWZ16 fit, as visible in Figure~\ref{fig:uBAR} (right), but it is more in line with what is predicted by the global PDF analyses.

\end{document}